\documentclass[conference]{IEEEtran}
\IEEEoverridecommandlockouts
\usepackage{cite}

\usepackage{amsmath,amssymb,amsfonts}
\usepackage[linesnumbered, ruled, vlined]{algorithm2e}
\usepackage{subcaption}
\usepackage{float}
\usepackage{algpseudocode}
\usepackage{graphicx}
\usepackage{cleveref}
\usepackage{textcomp}
\usepackage{xcolor}
\def\BibTeX{{\rm B\kern-.05em{\sc i\kern-.025em b}\kern-.08em T\kern-.1667em\lower.7ex\hbox{E}\kern-.125emX}}

\begin{document}
\title{Reputation-Driven Peer-to-Peer Live Streaming Architecture for Preventing Free-Riding}

\author{\IEEEauthorblockN{Rashmi Kushwaha, Rahul Bhattacharyya, Yatindra Nath Singh}
\IEEEauthorblockA{
\textit{Department of Electrical Engineering} \\
\textit{Indian Institute of Technology, Kanpur}\\
Kanpur, India \\
rashmikushwaha221@gmail.com, rahulbhatta0@gmail.com, ynsingh@iitk.ac.in}
}

\maketitle

\begin{abstract}
We present a peer-to-peer (P2P) live-streaming architecture designed to address challenges such as free-riding, malicious peers, churn, and network instability through the integration of a reputation system. The proposed algorithm incentivizes active peer participation while discouraging opportunistic behaviors, with a reputation mechanism that rewards altruistic peers and penalizes free riders and malicious actors. To manage peer dynamics, the algorithm continuously updates the strategies and adjusts to changing neighbors. It also implements a request-to-join mechanism for flash crowd scenarios, allowing the source node to delegate requests to child nodes, forming an interconnected tree structure that efficiently handles high demand and maintains system stability. The decentralized reputation mechanism promotes long-term sustainability in the P2P live streaming system. 
\end{abstract}


\section{Introduction}
Video content has become one of the largest and fastest-growing segments of internet traffic, driven by the popularity of streaming services, social media, and video conferencing platforms like YouTube, Netflix, and Instagram. Cisco's Visual Networking Index (VNI) \cite{b1}, projects that video will continue to dominate internet data consumption, encompassing everything from HD and UHD streaming to live broadcasts and user-generated content. The surge in video traffic is largely attributed to the increasing demand for high-definition content, live streaming, and the growing importance of remote work and online learning, especially accelerated by the COVID-19 pandemic.

Moreover, the rise of smartphones and other connected devices has further facilitated the accessibility of video content globally. Major global events, such as sports tournaments and political elections, often lead to spikes in video consumption as users seek live coverage. Additionally, the proliferation of security cameras and IoT devices has contributed to the increased demand for data bandwidth, generating more video content than ever before. Video traffic has thus become central to modern digital life, influencing how people communicate, learn, and interact online.

Presently, client-server architecture is the mainstay of live streaming paradigms, enabling content delivery from a central server to multiple clients. However, this approach faces significant challenges, particularly in terms of scalability and central point of failure, as the demand for server resources continues to grow. The centralized nature of this architecture makes it vulnerable to server overloads and outages, which can disrupt the streaming experience for all users.

In contrast, the P2P live streaming system presents a decentralized alternative that mitigates these issues. In a P2P setup, each peer functions as both a client and a server, redistributing the workload across the network with the help of DHT algorithms \cite{b26}. This collaborative sharing of resources reduces the strain on centralized servers, enhancing scalability and resilience. The decentralized nature of P2P streaming also improves fault tolerance, as the system is not dependent on a single point of failure. Additionally, the distributed content delivery model can reduce latency and buffering, providing a better streaming experience, especially for users spread across different locations.

While P2P live streaming offers advantages like scalability and resilience, it also faces challenges, particularly due to the behavior of selfish peers \cite{b2,b3,b4,b5,b6,b7,b8}, who consume resources without contributing. This can create imbalances in the network, leading to a degraded streaming experience for others. To address this, an incentive mechanism is essential to encourage active participation and cooperation among peers, ensuring a sustainable and efficient network.

One approach is token-based economies \cite{b13,b24,b25}, where peers are rewarded with tokens or cryptocurrency for contributing resources, which they can exchange for goods or services within the ecosystem. While effective, this method introduces complexities related to monetary transactions. Another approach is reputation-based incentives \cite{b9,b10}, where peers earn reputation points based on their contributions, with high-reputation peers receiving privileges and low-reputation peers facing penalties. This method fosters cooperation without the need for direct financial rewards, making it more adaptable to different network values.

By implementing reputation-based incentives, we seek to create a live streaming ecosystem where peers are motivated to contribute positively to the network, thereby enhancing its overall performance, reliability, and sustainability.

\par In this paper,
\begin{itemize}
\item We have introduced a reputation-based live streaming architecture consisting of five key stages. A reputation aggregation formula was defined to track peers' contributions.
\item We have established a game-theoretic equilibrium, where the system stabilizes based on peer satisfaction.
\end{itemize}
\par In this paper, section II presents the related work and section III presents our architecture for reputation-based live streaming system. Section IV presents Nash Equilibrium for system's stability. Finally conclusions are presented.

\section{Related work}
P2P live streaming offers a scalable, cost-effective alternative to traditional client-server models by utilizing the collective resources of peers to consume and distribute content. However, free-riding behavior \cite{b2,b3,b4,b5,b6,b7,b8}, where peers consume without contributing, challenges the system's effectiveness. To improve performance, reliability, and fairness, various incentive mechanisms have been developed to counteract free-riding.

The evolution of incentive mechanisms in P2P live streaming has led to a diverse range of approaches, including reputation-based systems \cite{b9, b10}, game-theoretic models \cite{b11, b12}, credit-based mechanisms \cite{b13,b24,b25}, and hybrid solutions \cite{b14}. 

Many authors in the past have contributed in this field. In \cite{b15}, the authors introduced PULSE, a P2P live streaming system that enhances video delivery through adaptive and incentive-based methods. It encourages cooperation by rewarding active peers with better streaming quality. But ensuring trust and preventing malicious behavior in an unstructured P2P network is challenging.

In \cite{b16}, the authors investigate strategies to promote cooperation in live P2P streaming by modeling the environment as a social network and using game theory to design incentive-based schemes. These schemes, involving reputation-based rewards and penalties, aim to encourage resource sharing and deter free-riding. While the strategies assume peer behavior can be influenced by incentives, real-world scenarios present unpredictability, with some peers acting selfishly despite these measures. In \cite{b17}, the authors introduced CoolStreaming, a data-driven overlay network where peers exchange data availability with a small set of partners and retrieve data from each other. Unlike tree-based systems, it offers better resilience to peer churn and adaptability through a pull-based data scheduling algorithm. However, this pull-based delivery can introduce latency, which may pose challenges for live streaming applications requiring low latency.

SplitStream \cite{b18}, is a content distribution system that uses multiple distribution trees to evenly distribute load in P2P networks. Each node acts as an interior node in one tree and a leaf in others, balancing network load. Built on Pastry, SplitStream provides higher bandwidth and robustness with low delay and overhead compared to single-tree multicast systems. However, managing multiple trees adds complexity to tree construction and maintenance. ChunkySpread \cite{b19}, is a P2P multicast system using a multi-tree framework to address tree- and mesh-based approach limitations. It handles peer capacity heterogeneity with a gossip-based protocol, ensuring balanced load distribution and fast adaptation to network changes. While it features loop avoidance and efficient data dissemination, managing unstructured trees in large networks can introduce significant overhead, affecting scalability.

PRIME (Peer-to-Peer Receiver-drIven MEsh-based Streaming) \cite{b20}, is a P2P live video streaming system using a mesh-based overlay and pull-based content delivery. Peers independently request content blocks from neighbors, improving adaptation to network dynamics and peer heterogeneity. However, the pull-based delivery can introduce latency, potentially impacting the real-time performance of live streaming. 

As P2P live streaming continues to grow and adapt to new challenges, the ongoing development \cite{b22,b23}, and refinement of incentive mechanisms remain essential for ensuring the long-term viability and success of the technology.

\section{System Architecture}
In a peer-to-peer (P2P) network, peers act as both clients and servers, sharing responsibility for system maintenance. Without a central authority, some peers may behave selfishly by downloading without contributing. To prevent this, our architecture implements a reputation-based incentive mechanism to reduce free-riding and encourage active participation.

Our architecture incorporates a Distributed Hash Table (DHT) layer that acts as a decentralized reputation management system. This DHT layer maintains and distributes the reputation records for all peers across the network. It enables efficient tracking of peer contributions, allowing for quick identification and accountability of network participants. 

We have considered a network of homogeneous peers, where each peer starts with the same reputation value (e.g., 0.5) and receives the same video quality, reputation is adjusted over time based on contributions. This evolving reputation influences peer connections, leading to a network topology shaped by the individual reputation scores of the peers.

\subsection{{Five key stages of architecture}}
Our system architecture can be structured into five key stages as follows:
\begin{itemize}
    \item Joining and authentication
    \item Initialization and topology formation
    \item Construction of routing tables
    \item Streaming of media
    \item Reputation calculation and aggregation
\end{itemize}
\vspace{1mm}
\subsubsection{\textbf{Joining and authentication}} Peers sign up with a username (email ID or mobile number) after downloading client software and join a common base DHT layer that supports various applications. The authentication manager (AM) verifies each user with an authentication server, generating a digitally signed certificate with the user's ID and public key. This allows secure mutual authentication, message signing, and encrypted communication between users using the SSL protocol, which ensures the security and integrity of the transactions.

When a peer initiates live streaming as a source, it creates a streamID and a stream descriptor with details like streamID, source nodeID, and content description, which is published with the bootstrap node. This streamID generates two DHT layers: one for media streaming and another for reputation tracking. There will be a mapping between streamID, sourceID, and corresponding DHT Layers for both media and reputation.

   \begin{figure}
    \centering
    \includegraphics[width=7cm]{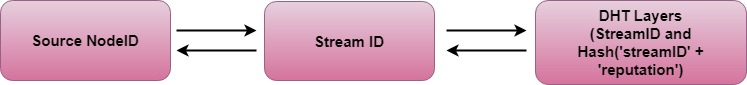}
    \setlength{\belowcaptionskip}{-8pt}
    \caption{Mapping of streamID with source ID and DHT layerIDs}
   \label{fig: Nash}
   \end{figure}
   Subscribing peers retrieve the streamID and the initial routing table (RT) from the bootstrap node, then update their RT through routing table management process and join/form the network for that stream. This ensures that they participate in both the media streaming and reputation DHT layers for the optimal performance.

\subsubsection{\textbf{Initialization and topology formation}} Peers will periodically update their RT-NT and establish a dedicated network for streaming purposes. When the source begins streaming, it initiates the transmission of beacons to all peers whose entry is there in its RT-NT.

Beacons are periodic message packets that maintain connectivity by indicating a node's readiness to provide media. Peers receiving these beacons can send "request-to-join" messages to potential parents in the BRT. If accepted, the requester is added as a child, and parent (grandparent) info is shared. If the parent is full, it provides its child node IDs as alternatives. This process forms an overlaid tree network for media distribution. Once connections are established, each peer maintains five tables based on the topology formed.

\subsubsection{\textbf{Construction of routing tables}} After a topology is established for each streamID, each peer will possess 5 tables. These tables assist the peers in sharing their media feed, staying connected during disruptions, and having backup peers in case of churn events.

\begin{enumerate}
\item \textbf{Routing Table (RT)}- This table will be created for each streamID (layerID), stores entries of participating nodes, with a maximum of 120 nodeIDs per our DHT algorithm. Initially, it is populated using the bootstrap node, and later maintained through the RT maintenance algorithm. Our architecture includes RT for both media streaming and the corresponding reputation layer.

\item \textbf{Neighbour Table (NT)}- This table will be initially created using RT by recording up to 16 nodeIDs closest to the node in RTT terms. In our architecture, NT is maintained only for media streaming and is periodically updated by exchanging RT-NT entries, retaining the lowest RTT(Round Trip Time) entries and discarding the rest.

\item \textbf{Broadcast Routing Table (BRT)}- This table will keep track of nodeIDs from which beacons are received and maintains entries for the top three nodes based on beacon consistency. This ensures system resilience, offering redundancy if one node fails. Once populated, a join request is sent to the first parent node, with a fallback to the next listed node if no response is received within a set period.

\item \textbf{Overlaid Multicast Table (OMT)}- This table will keep track of the parent peer from whom the media feed is received and the child peers to whom the feed is sent. While there is only one parent, multiple child nodes can receive the stream. Based on the BRT, join requests are sent sequentially to parent nodes, with the OMT recording the parent that accepts the request. If no response is received before the timer expires, the request is resent. Child nodes are removed if they fail to send a join request within the set time, while the timer resets upon receiving one.

\item \textbf{Backup Parents Table (BPT)}- This table will store information on grandparents and siblings to support network connectivity during disruptions. When a peer accepts a join request, it shares its parent (grandparent) info; if it rejects a request, it shares its children's info (potential siblings). The requesting peer stores this data in the BPT, aiding efficient recovery during churn or network issues.
\end{enumerate} 

\subsubsection{\textbf{Streaming of media}}
A peer in the network can function either as a source or a subscriber/ forwarder. Based on the role that the peer assumes, it will select either algorithm 1 or algorithm 2 accordingly.

\begin{algorithm}
\caption{Streaming Setup and Management for Source}
\begin{algorithmic}[1]
\State \textbf{Start}
\State {Enter details of streaming session(title, speaker name, date, and time)}
\State \textbf{Initialize:}
\State \hspace{1em} \textbf{a.} Create blank RT and NT
\State \hspace{1em} \textbf{b.} Register LAYERID = HASH(TITLE, NAME, DATE, TIME) with bootstrap node
\State \hspace{1em} \textbf{c.} Register Rep-LAYERID = HASH(layerID and repuation) with bootstrap node
\State \hspace{1em} \textbf{d.} Send RT and NT to bootstrap for layerID and RT for Rep-layerID,
\State \hspace{1em} \textbf{e.} Bootstrap starts maintaining RT and NT as leaf node in both LS and Rep-LS layer
\State \textbf{Start thread for creation and maintenance of routing tables and NT for LS layer, RT for Rep-LS}
\State \textbf{Start thread to periodically generate and send BEACONS() to all nodes listed in RT and NT of LS}
\State \textbf{Loop:}
\State \hspace{1em} Wait for JOINREQUESTS() from subscriber nodes
\State \hspace{1em} Respond to JOINREQUESTS() according to reputation
\State \hspace{1em} CREATEOMT() or use existing OMT to add/replace them as children
\State \hspace{1em} Start or continue STREAMING()
\State \hspace{1em} If streaming is ongoing, go back to step 12
\State \textbf{End}
\end{algorithmic}
\end{algorithm}

\begin{algorithm}
\caption{Streaming Setup and Management for Source}
\begin{algorithmic}[2]
\State \textbf{Start}
\State Paste the Stream ID
\State Get the initial RT for LS layer and LS-Rep from boostrap node
\State Start RT and NT maintenance thread
\State Start thread for receiving beacons and then forwarding them, while maintaining BRT
\State Send joinRequest() to the node as per BRT to start receiving the Stream
\State Create OMT (Overlay Multicast Table) as per acceptance received for sent joinRequests() and then put received media packets in queue for player, also forward them to children as per OMT
\State \textbf{Loop:}
\State \hspace{1em} Wait for JOINREQUESTS() from other subscriber nodes
\State \hspace{1em} Respond to JOINREQUESTS() according to reputation
\State \hspace{1em} Update existing OMT to add/replace them as children
\State Send joinRequest to grandparents from BPT (if reputation is higher than current parent) and after acceptance from grandparents, update OMT
\State \textbf{End Stream when:}
\State \quad Subscriber leaves \textbf{OR}
\State \quad Source ends streaming by sending info to all children, parent and all RT and NT entries
\State \textbf{End}
\end{algorithmic}
\end{algorithm}

\subsubsection{\textbf{Reputation calculation and aggregation}} A separate layer stores and updates reputation linked to the streamID, with peers in the streaming layer generally also present in the reputation layer. In this layer, only an RT table is needed. New peers are assigned an initial reputation, and when a peer receives a request, it checks the requester's reputation from the DHT layer before deciding to accept or reject the request.

\textbf{Reputation aggregration formula} can be represented as;
\begin{equation}
    R(t)=\frac{R(t-\tau) e^{-\alpha \tau}+R_r}{1+R_r},
\end{equation}
\hspace{5mm} where, $R_r=$ \textit{Reputation of receiving peers who is sending reputation value}, $R(t-\tau)=$ \textit{Last estimated value of reputation at time \lq$(t-\tau)$\rq}, $\alpha=$ \textit{Decay factor (rate at which reputation value will decay or forgotten)}. \textit{The value of $R(t)\in$ [0,1]}.
\vspace{2mm}

Peers earn higher reputations by sending video to requesters with higher reputation scores. Thus, for a node, optimal strategy is to forward media streams to as many as possible higher-reputation nodes. 

\textbf{Analysis of optimum value of \lq{$\alpha$}\rq}- Suppose a peer behaves as a free rider, receiving the feed without forwarding it to other peers. So, from reputation aggregation formula 
    $$R(t)=\frac{R(t-\tau) e^{-\alpha \tau}+R_r}{1+R_r},$$
    where, value of $R_r=0$ because no one is sending reputation update for the node (as there is no child), the formula becomes 
    $$R(t)={R(t-\tau) e^{-\alpha \tau}}.$$

\begin{figure}
    \centering
    \includegraphics[width=7.5cm]{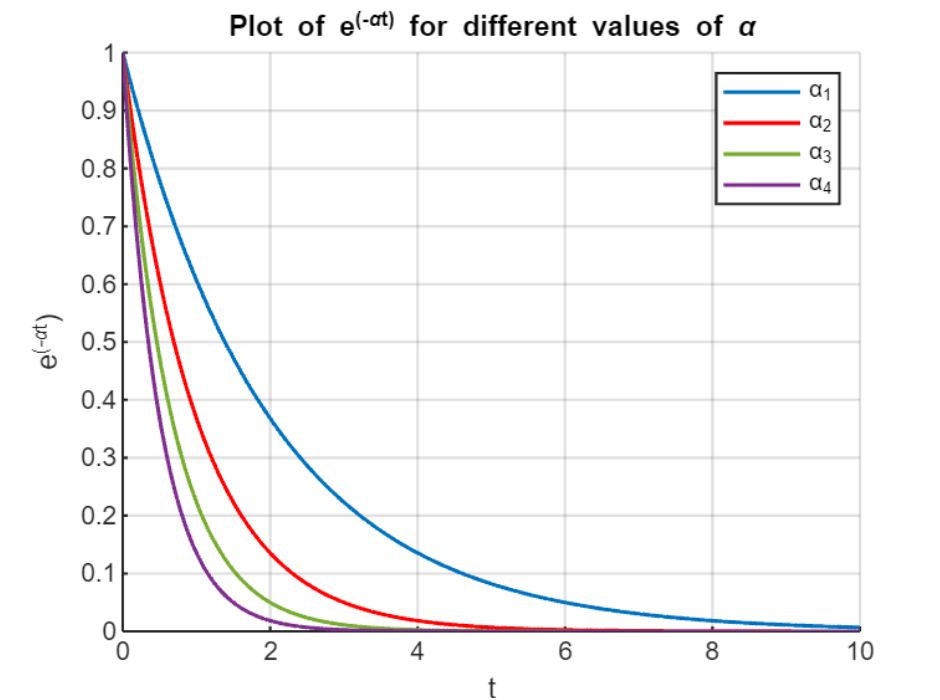}
    \setlength{\belowcaptionskip}{-10pt}
    \caption{Significance of \lq{$\alpha$}\rq}
   \label{fig: alpha}
   \end{figure}

Depending on the value of \lq{$\alpha$}\rq, the reputation of a free rider will gradually decay over time, leading to eventual pushing him to bottom of distribution tree and hence, he will get the poor quality and delayed media. Figure \ref{fig: alpha} shows the plot for decay of reputation for different values of alpha. Here, $\alpha_1<\alpha_2<\alpha_3<\alpha_4$, with higher value of \lq{$\alpha$}\rq, decay will be fast.
 \vspace{2mm} 
 
\textit{Argument: In live streaming, which is a continuous process with a finite duration, for maximizing system performance swiftly reducing the reputation of free riders is crucial. To achieve this, a larger value of {\lq{$\alpha$}\rq} is used to accelerate the decay of reputation. This ensures that free riders are swiftly identified and pushed to bottom of tree, thereby enhancing overall performance during the streaming.}

The reputation DHT layer keeps three identical copies of the reputation data distributed across different peers to prevent any single node from manipulating the reputation values. As a result, each peer typically stores the reputation values of three other peers on an average.

Once streaming starts, each node creates the necessary tables and begins supplying media feeds to its child peers. These child peers then send reputation updates to the root nodes of their feeding peers in the reputation DHT. The root nodes update the reputation of the feeder peers based on the received information. Peers use these updated reputation values to decide whether to accept or reject requests, favoring those with higher reputations. This process can unfold in two scenarios:

\begin{itemize}
    \item When peer accepts a join request, he will send his parents information to the children node (i.e. grandparents info). Figure \ref{fig:NiceImage}\subref{fig:NiceImage1} shows this scenario. Here, node $C$ will tell about node $A$ to nodes $E$, $F$ and $G$.

  \begin{figure}
    \centering
      \begin{subfigure}{0.2\textwidth}
        \includegraphics[width=\textwidth]{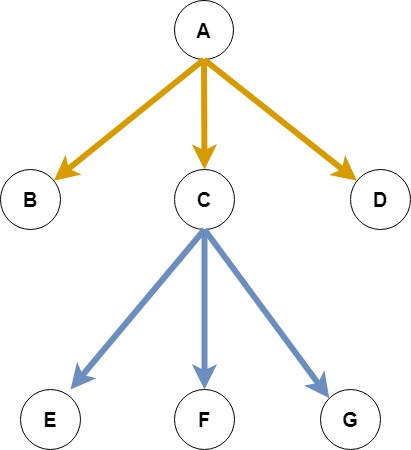}
          \caption{}
          \label{fig:NiceImage1}
      \end{subfigure}
      \hfill
      \begin{subfigure}{0.25\textwidth}
        \includegraphics[width=\textwidth]{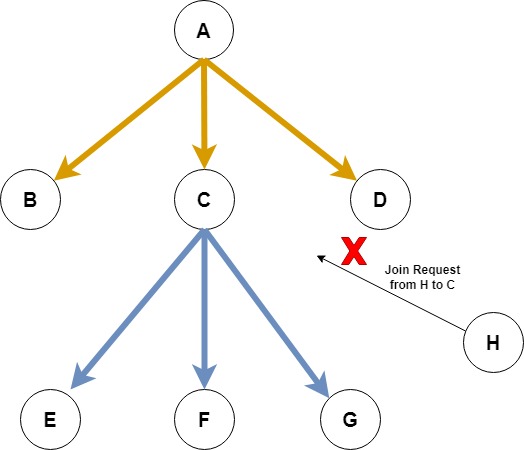}
          \caption{}
          \label{fig:NiceImage2}
      \end{subfigure}
      \setlength{\belowcaptionskip}{-10pt}
\caption{
\label{fig:NiceImage}%
Topology showing acceptance and rejection of peers's request}
\end{figure}

\item When the peer rejects the join request, he will send his children nodeIDs to the peer, which can be used to send join request. Figure \ref{fig:NiceImage}\subref{fig:NiceImage2} shows this scenario. Here, node $H$ has lower reputation than nodes $E$, $F$ and $G$ hence, node $C$ will reject $H$'s request and tell him about his children($E$, $F$ and $G$). $H$ in turn can now try sending a join request to either of $E$, $F$ and $G$.
\end{itemize}

\begin{figure}
    \centering
    \includegraphics[width=6cm]{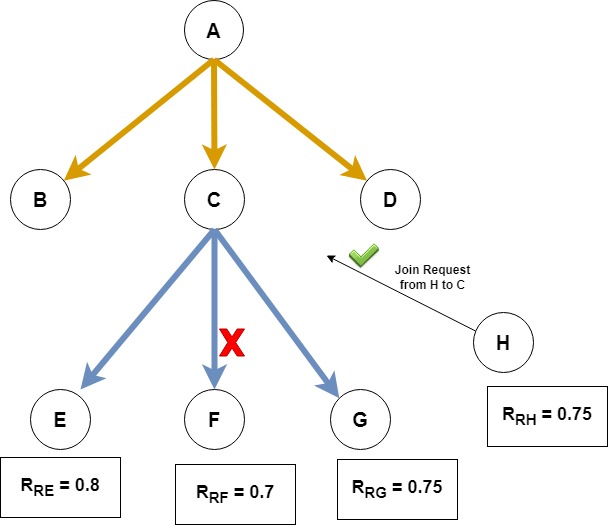}
    \setlength{\belowcaptionskip}{-10pt}
    \caption{Topology}  
   \label{fig: T3}
\end{figure}

During streaming, if a peer receives a request from another peer with a higher reputation than any one of its current child nodes, it will replace the lower-reputation child in its OMT with the higher-reputation peer. Before replacing the lower-reputation child, it will provide the evicted peer with the node IDs of its other child nodes, which can be used by the evicted node as feeder. As depicted in figure \ref{fig: T3}.

Suppose that $H$ has a higher reputation than $F$, so $C$ will replace $F$ with $H$, but before replacing, $C$ will tell $F$ about his children's node ($E$, $G$ and $H$).

The peers have grandparents info with them, so they will try to connect with grandparents to increase their video quality and reducing delay. Additionally, if a peer is streaming to 10 other peers, it can check the reputation DHT to verify whether each of those peers is submitting their reputation updates and whether the updates are accurate. This ensures accountability and correctness in reputation reporting.

\section{{Stability of the system}}
\hspace{5mm}We have defined the system's stability using Nash equilibrium in game theory. There are two types of peers in the system
\begin{itemize}
    \item Altruistic peers (Good peers)
    \item Free riders and Malicious peers (Bad peers)
\end{itemize}

\textit{Assumption: Majority of peers are altruistic}.

\textbf{In case of good peers-} Altruistic peers will report the true value of reputation and serve according to their own capacity. They do not need to devise any strategies, as they simply follow the rules implemented for the system to work effectively.

\textbf{In case of bad peers-} 
\begin{enumerate}
    \item Free riders- When encountering an altruistic peer, they will have their true reputation value reported to other peers, allowing them to be easily detected based on their reputation. 
    When encountering bad peers, they may either report their true value or provide misleading information. However, our system employs an aggregation formula that takes into consideration the majority of altruistic peers who provide accurate reputation reports. 
\textit{Malicious collective will not play any role.}

\item Malicious peer- When encountering altruistic peers, malicious peers send inaccurate reputation values, but the majority aggregation mechanism plays a crucial role. Furthermore, our system allows the sending peer to inspect the reputation value received from the receiving peer. 
\end{enumerate}  
    
\begin{figure}
    \centering
    \includegraphics[width=04cm]{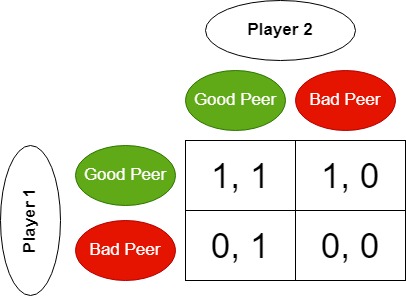}
     \setlength{\belowcaptionskip}{-10pt}
    \caption{Payoff Matrix}
    \label{fig: Payoff Matrix}
   \end{figure}

Now, according to the above scenarios, if we draw a payoff matrix then, we have figure \ref{fig: Payoff Matrix}.

\textbf{1}- Their reputation value contributes to the aggregation process, meaning that the reputation they send is taken into account based on the majority consensus.

\textbf{0}- Their reputation value is ignored in the aggregation process, meaning that the reputation they send does not impact the overall aggregation.

\textbf{What equilibrium can be achieved in the system?} At equilibrium, if a peer receives a higher payoff, they will not change their strategy. This means that deviating from the equilibrium would result in a loss for the peer.

\textbf{Altruistic Peers}- They gain because they are actively serving, ensuring their reputation remains high. Additionally, their accurate reporting of others' reputations ensures their contribution is considered in the aggregation process.

\textbf{Free riders}- They incur loss because they do not serve. Consequently, their reputation decays over time, leading to eventual expulsion from the system.

\textbf{Malicious peers}- They are also at a disadvantage because the reputation they send does not influence the aggregation process. Their malicious intent fails due to the majority of honest contributions.

   \begin{figure}[H]
    \centering
    \includegraphics[width=04cm]{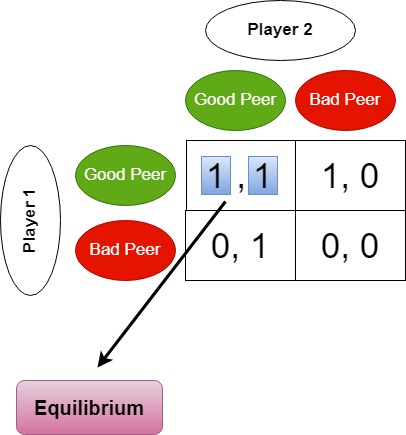}
    \setlength{\belowcaptionskip}{-9pt}
    \caption{Nash equilibrium}
    \label{fig: Nash equilibrium}
   \end{figure}
   
Figure \ref{fig: Nash equilibrium}, shows the Nash equilibrium. From the statements above, we observe that, at equilibrium, both free riders and malicious peers will be removed from the system, while good peers will receive higher payoffs to contribute more. As a result, good peers' reputation will increase (up to 1), and their demands will be met. With satisfied altruistic peers, the system's performance will ultimately enhance in terms of streaming quality and minimal delay, crucial for live streaming. The system will be stabilize in terms of peers' satisfaction.

\section{Conclusion}
Our algorithm effectively addresses critical challenges in peer-to-peer live streaming, such as free-riding, malicious peers, churn, and flash crowds. By incorporating a robust reputation and incentive system, the algorithm fosters active participation and discourages selfish behaviors, ensuring a balanced network. Its dynamic adaptation to peer behavior and changing network conditions allows for continuous optimization of strategies and neighbor relationships. For flash crowd scenarios, the request-to-join mechanism efficiently manages high demand, forming an interconnected tree structure that enhances system stability. The reputation mechanism is central to this architecture, providing clear incentives for positive contributions, improving overall performance, and promoting the long-term sustainability of the P2P live streaming system. 

\textbf{\textit{Limitation of this algorithm}}: Each peer receives the same video quality regardless of their contribution level to the system. Additionally, this algorithm does not consider variations in peer's bandwidth.


\begin{thebibliography}{00}
\bibitem{b1}Cisco(2020). Cisco Annual Internet Report (2018–2023) White Paper. Retrieved from https://www.cisco.com/c/en/us/solutions/collateral/executive perspectives/annual-internet report/white-paper-c11-741490.html.

\bibitem{b2}E. Adar, and BA. Huberman, “Free riding on gnutella," (2000).

\bibitem{b3}S. Saroiu, PK. Gummadi, and SD. Gribble, “Measurement study of peer-to-peer file sharing systems," In \textit{Multimedia computing and networking 2002}, vol. 4673, pp. 156-170, SPIE, 2001.

\bibitem{b4}D. Hughes, G. Coulson, and J. Walkerdine, “Free riding on Gnutella revisited: the bell tolls?," \textit{IEEE distributed systems online 6}, no. 6 2005.

\bibitem{b5}J. Liang, R. Kumar, and KW. Ross, “The FastTrack overlay: A measurement study," \textit{Computer Networks 50}, no. 6: 842-858, 2006.


\bibitem{b6}JJD. Mol, JA. Pouwelse, M. Meulpolder, DHJ. Epema, and HJ. Sips. “Give-to-get: free-riding resilient video-on-demand in p2p systems," In \textit{Multimedia Computing and Networking 2008}, vol. 6818, pp. 27-34. SPIE, 2008.

\bibitem{b7}S. Kaune, RC. Rumin, G. Tyson, A. Mauthe, C. Guerrero, and R. Steinmetz, “Unraveling bittorrent's file unavailability: Measurements and analysis," In \textit{2010 IEEE Tenth International Conference on Peer-to-Peer Computing (P2P)}, pp. 1-9. IEEE, 2010.

\bibitem{b8}N. Andrade, M. Mowbray, A. Lima, G. Wagner, and M. Ripeanu, “Influences on cooperation in bittorrent communities," In \textit{Proceedings of the 2005 ACM SIGCOMM workshop on Economics of peer-to-peer systems}, pp. 111-115. 2005.

\bibitem{b9}SD. Kamvar, MT. Schlosser, and H. Garcia-Molina, “The eigentrust algorithm for reputation management in p2p networks," In \textit{Proceedings of the 12th international conference on World Wide Web}, pp. 640-651, 2003.

\bibitem{b10}R. Zhou, and K. Hwang, “Powertrust: A robust and scalable reputation system for trusted peer-to-peer computing," \textit{IEEE Transactions on parallel and distributed systems 18}, no. 4: 460-473, 2007.

\bibitem{b11}X. Kang, and Y Wu, “A game-theoretic approach for cooperation stimulation in peer-to-peer streaming networks," In \textit{2013 IEEE International Conference on Communications (ICC)}, pp. 2283-2287, IEEE, 2013.

\bibitem{b12}J. Park, and M Van der Schaar, “A game theoretic analysis of incentives in content production and sharing over peer-to-peer networks," \textit{IEEE journal of selected topics in signal processing 4}, no. 4: 704-717, 2010.

\bibitem{b13}TWJ. Ngan, DS. Wallach, and P. Druschel, “Enforcing fair sharing of peer-to-peer resources," In \textit{Peer-to-Peer Systems II: Second International Workshop, IPTPS 2003, Berkeley, CA, USA, February 21-22}, Revised Papers 2, pp. 149-159, Springer Berlin Heidelberg, 2003.

\bibitem{b14}M. Feldman, L. Kevin, I Stoica, and J Chuang, “Robust incentive techniques for peer-to-peer networks," In \textit{Proceedings of the 5th ACM conference on Electronic commerce}, pp. 102-111, 2004.

\bibitem{b15}F. Pianese, D. Perino, K. Joaquín, and EW. Biersack, “PULSE: an adaptive, incentive-based, unstructured P2P live streaming system," \textit{IEEE Transactions on Multimedia 9, no. 8}: 1645-1660, 2007.

\bibitem{b16}WS. Lin, HV. Zhao, and KJR. Liu, “Incentive cooperation strategies for peer-to-peer live multimedia streaming social networks," \textit{IEEE transactions on multimedia 11}, no. 3: 396-412, 2009.

\bibitem{b17}X. Zhang, J. Liu, B. Li, and YSP. Yum, “CoolStreaming/DONet: A data-driven overlay network for peer-to-peer live media streaming," In \textit{Proceedings IEEE 24th Annual Joint Conference of the IEEE Computer and Communications Societies.}, vol. 3, pp. 2102-2111, IEEE, 2005.

\bibitem{b18}M. Castro, P. Druschel, AM. Kermarrec, A. Nandi, A. Rowstron, and A. Singh, “Splitstream: High-bandwidth multicast in cooperative environments," \textit{ACM SIGOPS operating systems review 37}, no. 5: 298-313, 2003.

\bibitem{b19}V. Venkataraman, K. Yoshida, and P. Francis, “Chunkyspread: Heterogeneous unstructured tree-based peer-to-peer multicast," In \textit{Proceedings of the 2006 IEEE international conference on network protocols}, pp. 2-11. IEEE, 2006.

\bibitem{b20}N. Magharei, and R. Rejaie, “Prime: Peer-to-peer receiver-driven mesh-based streaming," \textit{IEEE/ACM transactions on networking 17}, no. 4: 1052-1065, 2009.

\bibitem{b21}L. Liu, J. Zou, and H. Xiong, “Bandwidth allocation for video streaming over peer-to-peer networks with nash bargaining," In \textit{Proceedings of International Conference on Internet Multimedia Computing and Service}, pp. 363-367. 2014.

\bibitem{b22}D. Halder, P. Kumar, S. Bhushan, and AM. Baswade, “FybrrStream: A WebRTC based efficient and scalable P2P live streaming platform," In \textit{2021 International Conference on Computer Communications and Networks (ICCCN)}, pp. 1-9, IEEE, 2021.

\bibitem{b23}SA. Ansari, K. Pal, MC. Govil, M. Ahmed, T. Chawla, and A. Choudhary, “Score-based Incentive Mechanism (SIM) for live multimedia streaming in peer-to-peer network," \textit{Multimedia Tools and Applications 80}: 19263-19290, 2021.

\bibitem{b24}M. Hazazi, A. Almousa, H. Kurdi, S. Al-Megren, and S. Alsalamah, “A Credit-Based Approach for Overcoming Free-Riding Behaviour in Peer-to-Peer Networks," \textit{Computers, Materials \& Continua 59}, no. 1, 2019.

\bibitem{b25}Z. ImaniMehr, and M. DehghanTakhtFooladi, “Token-based incentive mechanism for peer-to-peer video streaming networks," \textit{The Journal of Supercomputing 75}: 6612-6631, 2019.

\bibitem{b26}R. Kushwaha, S. Kulkarni, and YN. Singh, “Generalized distance metric for different DHT routing algorithms in peer-to-peer networks," \textit{arXiv preprint arXiv:2303.13965}, 2023.

\end{thebibliography}
\end{document}